\documentclass[aip,jcp,reprint,amsmath,amssymb]{revtex4-1}

\usepackage{graphicx}
\usepackage{dcolumn}
\usepackage{bm}

\usepackage[utf8]{inputenc}
\usepackage[T1]{fontenc}
\usepackage{mathptmx}
\usepackage{etoolbox}

\makeatletter
\def\@email#1#2{%
 \endgroup
 \patchcmd{\titleblock@produce}
  {\frontmatter@RRAPformat}
  {\frontmatter@RRAPformat{\produce@RRAP{*#1\href{mailto:#2}{#2}}}\frontmatter@RRAPformat}
  {}{}
}%
\makeatother
\begin{document}

\preprint{AIP/123-QED}

\title{A General Molecular-Scale Dynamic Memristor Model Based on Non-equilibrium Charge Transport Kinetics and Its Information Processing Capability in Reservoir Computing }
\author{Yueqi Chen}
\affiliation{%
State Key Laboratory of Advanced Materials for Intelligent Sensing, Key Laboratory of Organic Integrated Circuit, Ministry of Education \& Tianjin Key Laboratory of Molecular Optoelectronic Sciences, Department of Chemistry, School of Science, Tianjin University, Tianjin 300072, China
}%
\affiliation{ 
Collaborative Innovation Center of Chemical Science and Engineering (Tianjin), Tianjin 300072, China
}%
\author{Xuan Ji}%
\email{jixuan\_0808@tju.edu.cn}
\affiliation{%
State Key Laboratory of Advanced Materials for Intelligent Sensing, Key Laboratory of Organic Integrated Circuit, Ministry of Education \& Tianjin Key Laboratory of Molecular Optoelectronic Sciences, Department of Chemistry, School of Science, Tianjin University, Tianjin 300072, China
}%
\affiliation{ 
Collaborative Innovation Center of Chemical Science and Engineering (Tianjin), Tianjin 300072, China
}%

\author{Xi Yu}
\email{xi.yu@tju.edu.cn}
\affiliation{%
State Key Laboratory of Advanced Materials for Intelligent Sensing, Key Laboratory of Organic Integrated Circuit, Ministry of Education \& Tianjin Key Laboratory of Molecular Optoelectronic Sciences, Department of Chemistry, School of Science, Tianjin University, Tianjin 300072, China
}%
\affiliation{ 
Collaborative Innovation Center of Chemical Science and Engineering (Tianjin), Tianjin 300072, China
}%

\date{\today}

\begin{abstract}
Non-equilibrium molecular-scale dynamics, where fast electron transport couples with slow chemical state evolution, underpins the complex behaviors of molecular memristors, yet a general model linking these dynamics to neuromorphic computing remains elusive. We introduce a dynamic memristor model that integrates Landauer and Marcus electron transport theories with the kinetics of slow processes, such as proton/ion migration or conformational changes. This framework reproduces experimental conductance hysteresis and emulates synaptic functions like short-term plasticity (STP) and spike-timing-dependent plasticity (STDP). By incorporating the model into a reservoir computing (RC) architecture, we show that computational performance optimizes when input frequency and bias mapping range align with the molecular system’s intrinsic kinetics. This chemistry-centric, bottom-up approach provides a theoretical foundation for molecular-scale neuromorphic computing, demonstrating how non-equilibrium molecular-scale dynamics can drive information processing in the post-Moore era.  
\end{abstract}

\maketitle

\section{\label{sec:level1a}Introduction}
Non-equilibrium molecular-scale dynamics, where fast electron transport couples with slow chemical or structural processes, gives rise to rich, history-dependent behaviors that are central to molecular-scale science. Processes such as proton/ion migration\cite{1,2,3}, conformational switching\cite{4,5}, or local charge accumulation\cite{6,7,8,9} introduce disparate timescales, leading to nonlinear phenomena like memristance. This emergent behavior is aligned with the theory of the memristor, first conceptualized by Chua\cite{10,11}, and later physically realized by HP Labs\cite{12} , whose unique ability to couple conductance with memory states makes the memristor ideal for artificial synapses in neuromorphic computing\cite{13}. Memristor dynamics are typically described by coupled equations, where a slow-evolving internal state variable (e.g., ionic migration) modulates fast electronic transport, governing conductance,\cite{12}
\begin{equation}
I=G(w, V) V
\label{eq:1},
\end{equation}
\begin{equation}
\frac{d w}{d t}=f(w, V)
\label{eq:2},
\end{equation}
where $w$ is the internal state variable of the device that evolves over time under the influence of an external voltage $V$, and the time-dependent evolution of $w$ governs the conductance $G$. This general dynamic description has been widely validated in solid state systems\cite{13,14,15,16}, which have been applied in nonvolatile memory\cite{17,18,19}, artificial synapses for neuromorphic computing\cite{13,20,21,22,23}, and information processing\cite{24,25,26}.

This class of memristive behavior has been actively explored in molecular junctions. Pioneering studies, such as theoretical studies by Nitzan et al. first proposed a non-steady-state transport model based on coupling fast tunneling with slow hopping transport in chemistry system.\cite{27,28} While this foundational work was conceptual, it was later experimentally realized and extended by Nijhuis et al., who demonstrated that such mechanisms in molecular junctions give rise to pronounced history-dependent conductance and emergent behaviors analogous to synaptic functions, providing a key link to neuromorphic computing.\cite{2,29,30} Yan et al. successfully demonstrated synaptic-like behavior and neuromorphic computation capability by coupling metal ion migration with charge transport in peptide junction.\cite{31} Our own work on an anthraquinone system modeled its proton-coupled electron transfer dynamics, successfully capturing memristive properties, synaptic functions, and neuromorphic computing capabilities. \cite{3,32} While insightful, these studies are often system-specific, underscoring the urgent need for a more general theoretical framework. Crucially, a key unanswered question is how these complex, multi-timescale molecular-scale dynamics can be harnessed for effective information processing. Establishing a general principle that links a device's intrinsic kinetic properties to its computational performance remains a central challenge in the field.
\begin{figure*}
    \centering
    \includegraphics[width=0.75\linewidth]{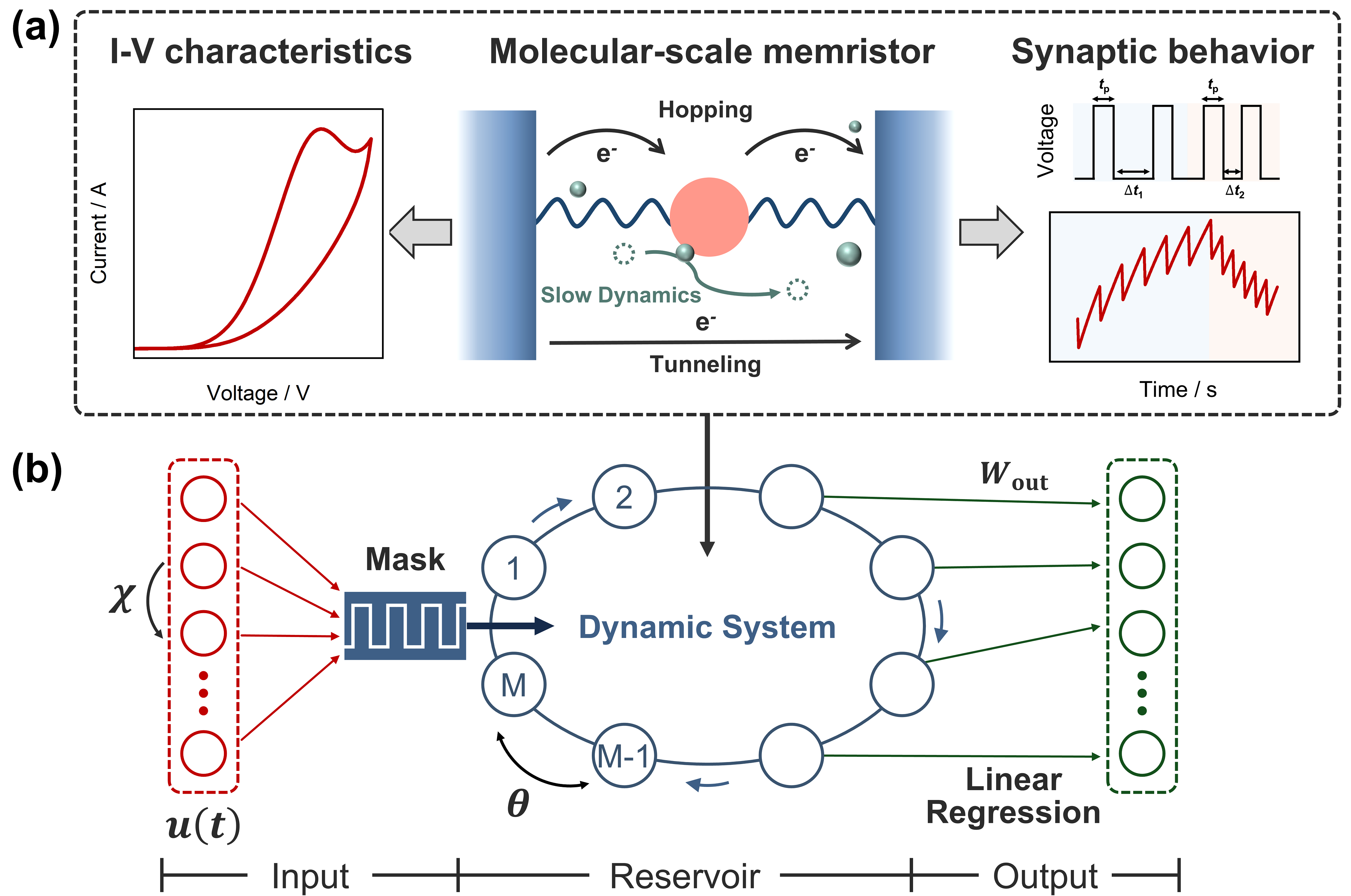}
    \caption{Schematic of molecular-scale dynamic memristors for neuromorphic computing. (a) Fast charge transport via hopping or tunneling, coupled with slow dynamics (middle), produces memristive hysteresis in the current-voltage (I-V) curve (left) and enables the emulation of synaptic plasticity (right). (b) Reservoir computing in a time-multiplexing architecture embeds dynamic memristors as the reservoir, efficiently exploiting their intrinsic dynamics for information processing. The input signal $u(t)$ is multiplexed by a mask and fed into a cyclic reservoir containing $M$ virtual nodes. The output weight $W_{\mathrm{Out}}$ is trained using linear regression to map the high-dimensional reservoir states to the desired output.
}
    \label{fig:1}
\end{figure*}

To address this gap, we develop a general dynamic memristor model for molecular system, coupling fast electron transport—described by Landauer and Marcus theories\cite{33,34}—with slow chemical state evolution (FIG.~\ref{fig:1}a). This bottom-up, chemistry-centric model provides a quantitative basis for molecular memristors, explaining how bias-induced changes in molecular states govern conductance. It successfully explains dynamic features like conductance hysteresis and reveals the molecular origins of emergent functional behaviors analogous to short-term plasticity (STP).\cite{35,36,37,38} and spike time dependent plasticity (STDP).\cite{39} To forge a quantitative link between the system's internal kinetics and its information processing capacity, we integrate our dynamic model into a reservoir computing (RC) framework\cite{40,41,42,43,44} (FIG.~\ref{fig:1}b). By tuning the two key hyperparameters in the RC—input frequency and bias mapping range—to align with the system's kinetic properties, we optimize performance for tasks like temporal pattern recognition and chaotic time-series prediction. This work thus establishes a bottom-up, theoretical foundation for understanding and harnessing molecular-scale dynamics in neuromorphic computing, offering design principles for the co-optimizing of molecular system and information processing paradigms.

\section{\label{sec:level1b}theory and model}

\subsection{\label{sec:level2a}The Dynamical Process}

\begin{figure}[!htbp]
    \centering
    \includegraphics[width=1\linewidth]{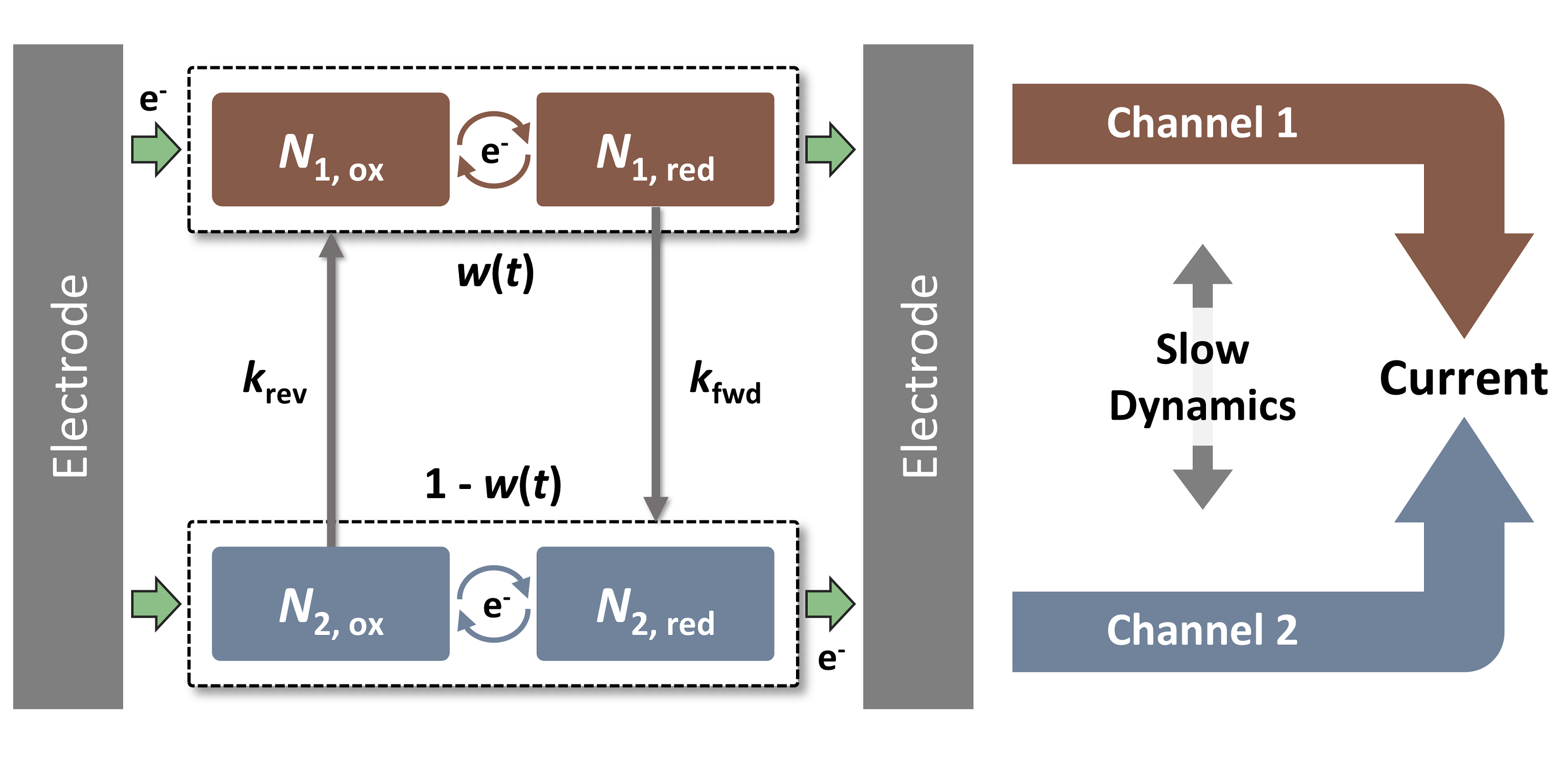}
    \caption{Schematic illustration of the molecular-scale dynamic memristor model that couples rapid charge transport with slow chemical dynamics. In each channel, steady-state charge transport proceeds via either hopping or tunneling, establishing an electron occupancy. This occupancy generates intermediate redox states that participate in the chemical dynamics, resulting in channel evolution and contributing to the overall current as a weight sum.
}
    \label{fig:2}
\end{figure}

In molecular dynamic memristive systems, the application of external bias induces rapid charge transport processes, during which the molecule gains or loses electrons, forming transient reduced or oxidized intermediate states.\cite{3,28} As shown in FIG.~\ref{fig:2}, these redox intermediates establish charge transport pathways that couple with the slow dynamic processes such as proton/ion migration, conformational switching, thereby enabling the emergence of a new charge transport channel. This dynamics, similar to the internal state dynamics in HP model, is modeled as a gradual transition from an initially low-conductance state (LCS, corresponding to Channel 1) to a high-conductance state (HCS, corresponding to Channel 2), with the latter also undergoing back-transition at a certain rate. The internal dynamics of the system is significantly lower than that of charge transport. Consequently, the system's inability to relax instantaneously endows it with a memory of its recent history, rendering its charge transport properties markedly time-dependent.\cite{28}

The dynamics is quantified by expressing the current as a weighted sum of the low- and high-conductance branches:
\begin{equation}
I_{\mathrm{Total}}(t)=w(t)I_{\mathrm{Low}}+\begin{pmatrix}1-w(t)\end{pmatrix}I_{\mathrm{High}}
\label{eq:3}.
\end{equation}
The normalized state variable $w(t)$denotes the fraction of the molecular-scale memristor in the LCS, while $1-w(t)$ is the complementary fraction in the HCS. The fraction of two states evolves following the first-order reaction kinetics.\cite{3},
\begin{equation}
\frac{dw}{dt}=-k_{\mathrm{fwd}}w+k_{\mathrm{rev}}(1-w)
\label{eq:4},
\end{equation}
where the forward and reverse reaction rates $k_{\mathrm{fwd/rev}}$ govern the slow internal transition between two channels, and being bias-dependent, are defined by:
\begin{equation}
k_{\mathrm{fwd}}(V)=N_{1,\mathrm{red}}(V)k_{+}\
\label{eq:5},
\end{equation}
\begin{equation}
k_{\mathrm{rev}}(V)=N_{2,\mathrm{ox}}(V)k_{-}
\label{eq:6}.
\end{equation}
Here, $k_{+}$ and $k_{-}$  are intrinsic rate constants that characterize the forward (LCS → HCS) and reverse (HCS → LCS) transitions, respectively. $N_{1,\mathrm{red}} $ and $N_{2,\mathrm{ox}}$  denote the electron and hole occupancy in Channel 1 and Channel 2 respectively, which determines the availability of intermediate redox states under bias. The values $k_{+}$=1000 and $k_{-}$=0.05 are deliberately chosen to create strong kinetic asymmetry, ensuring robust, hysteretic switching behavior. The transition rates are modulated by the bias-dependent occupancy $N_{1,\mathrm{red}} $ and $N_{2,\mathrm{ox}} $: the former enhances the forward switching by increasing the population of reduced state, while the latter facilitates the reverse transition by quantifying the availability of oxidized state, which is related to the reduced state by $N_{2,\mathrm{ox}}(V)=1-N_{2,\mathrm{red}}(V)$. Although $N_{1,\mathrm{red}}$ and $N_{2,\mathrm{red}}$ pertain to different channels, they follow an identical voltage-dependent distribution described by either the Landauer-Büttiker formalism or Marcus theory. The numerical implementation of the dynamic model is detailed in Section 1 of the Supplementary Material. 

\subsection{\label{sec:level2b}Steady-State Charge Transport in the Model}

Steady-state charge transport in molecular junctions proceeds with two main mechanisms: coherent tunneling, captured by the Landauer formalism\cite{34}, and incoherent hopping, described by Marcus theory\cite{33}(see Section 2 of the Supplementary Material). These pathways produce different bias-dependent populations of the oxidized and reduced states, and, from a dynamic perspective, impart distinct dynamical characteristics. To construct general dynamic models without extensive parameter fitting, we distilled their essential bias dependences into the following closed-form expressions: 
\begin{equation}
N_{1/2,\mathrm{red}}^{\mathrm{H}}=\cosh\left(5|V|\right)\cdot10^{-5}\
\label{eq:7},
\end{equation}
\begin{equation}
N_{1/2,\mathrm{red}}^{\mathrm{T}}=\cosh\left(1.2|V|^{1.2}\right)\cdot10^{-3.6}
\label{eq:8}.
\end{equation}
In the hopping transport, the electron occupancy exhibits an exponential dependence on bias.\cite{45} Rescaling the argument of the function by a factor of 5 reproduces the near-exponential increase in electron occupancy characteristic of activation-controlled hopping. In contract, for the tunneling transport, a smaller scaling factor together with an additional bias exponent of 1.2 produces a weaker dependence that is quasi-linear at low bias and becomes nonlinear near resonance at higher bias.\cite{46} The numerical prefactors ($10^{-5}$ and $10^{-3.6}$) keep the occupancy within physically reasonable bounds.

Although the two forms of transport differ in their bias-dependent kinetics, the steady-state current is described by the same analytical expression commonly used in molecular junction studies,\cite{47}
\begin{equation}
I_{\mathrm{High/Low}}=G_{\mathrm{High/Low}}\cdot\left(V+\left(\frac V{0.33}\right)^3\right)
\label{eq:9}.
\end{equation}
With $G_{\mathrm{High}} = 1.0 \cdot 10^{-10}$ and $G_{\mathrm{High}} = 5.0 \cdot 10^{-12}$, the 20-fold difference renders the transition dynamics clearly discernible. The specific current-voltage (I-V) relationship, however, is less critical. As shown in Section 7 of the Supplementary Material, the system retains its ability to effectively process information even when the I-V relationship is simplified from a cubic to a linear form. This indicates that the computational power is primarily governed by the dynamics that are dictated by the bias-dependent charge occupancy states, which validates our use of the simplified Eq.~(\ref{eq:9}) as a sound and economical choice for the generic model. In addition, to better mimic experimental conditions, we add Johnson–Nyquist thermal noise\cite{48} to the model current by:
\begin{equation}
I_{\text{Noise}} = \sqrt{\frac{4k_B T B_w}{R}}
\label{eq:10},
\end{equation}
where the noise bandwidth $B_w$ is set to $10^{4}$ to preserve a reasonable signal-to-noise ratio,\cite{49} $R=\frac{dV}{dI}$ is the differential resistance of each transport channel,  $k_B$ is the Boltzmann constant, and $T$ is the absolute temperature (300 K). 

\section{\label{sec:level1c}Result and discussion}
We investigate the dynamic properties of molecular memristors through the lens of non-equilibrium molecular-scale dynamics, including I-V hysteresis characteristics, synaptic emulation, and information processing by RC—primarily through the hopping transport. The molecular system exhibits characteristic hysteretic switching and can emulate key synaptic behaviors like STP and STDP in response to varying input frequencies. While the memristors with the hopping and tunneling transport produce qualitatively similar results for these time-dependent behaviors (see Supplementary Material for tunneling transport results), they differ significantly when assessing the impact of bias mapping range on reservoir computing. This is due to the distinct bias-dependence of electron occupancy in two types of transport (Eqs.~(\ref{eq:7})--(\ref{eq:8})), which in turn affects the selection of RC hyperparameters. Accordingly, we will provide a detailed comparison of the two types of transport specifically for the analysis of bias-dependent RC information processing.

\subsection{\label{sec:level2c}Emulation of Dynamic Hysteresis  }

\begin{figure}[!htbp]
    \centering
    \includegraphics[width=1\linewidth]{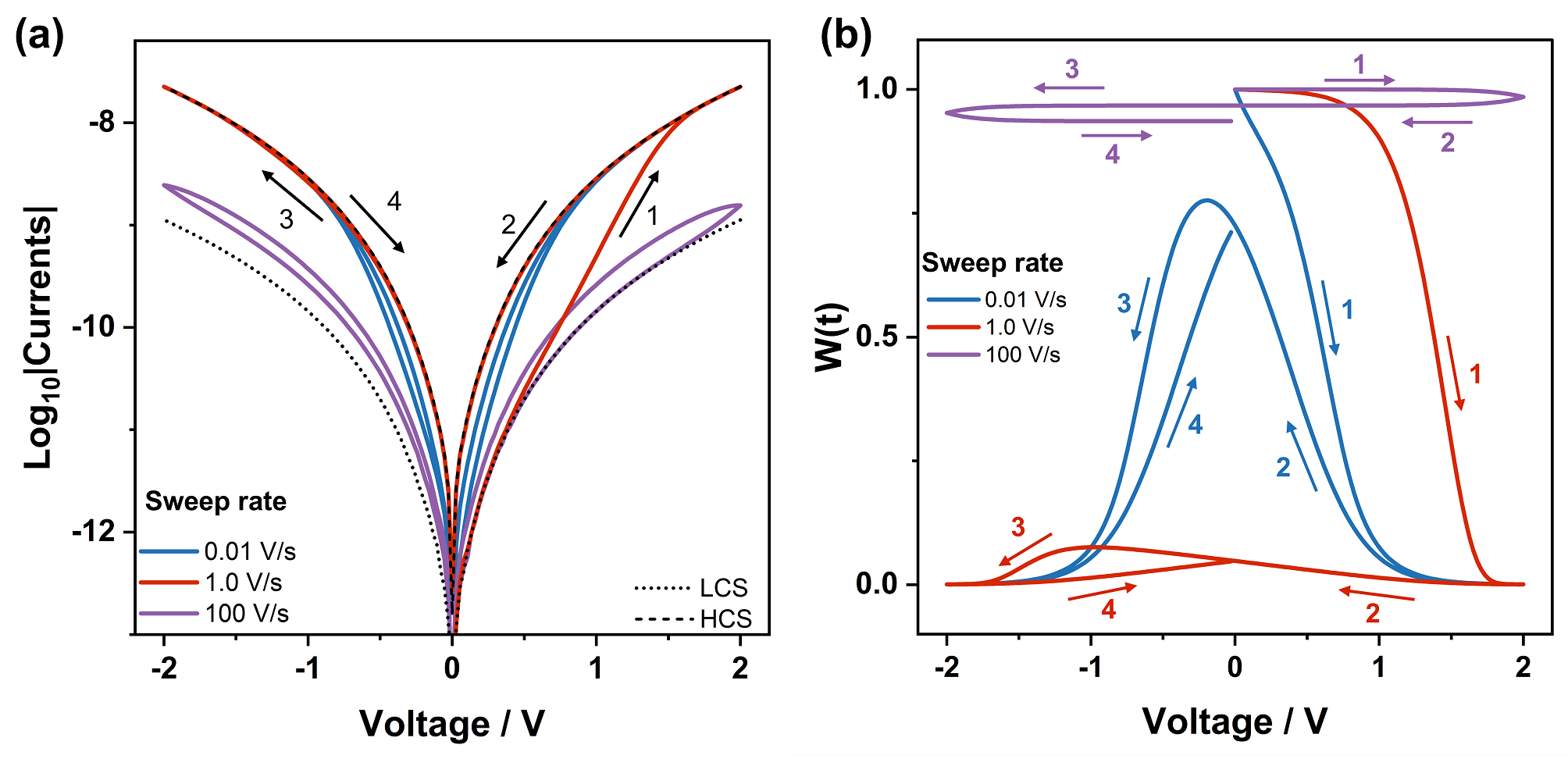}
    \caption{Dynamic I-V hysteresis and evolution of the LCS fraction in the hopping-based memristor. (a) I-V characteristics at slow, moderate, and fast voltage sweep rates, respectively, showing the dynamic current of the system alongside the two steady-state currents. (b) Evolution of the LCS fraction under slow, moderate, and fast sweep rates. Significant hysteresis and state evolution are observed only at the 1.0 V/s sweep rate.}
    \label{fig:3}
\end{figure}

Arising from the coupled molecular dynamics, our molecular-scale memristor model gives rise to a pinched hysteresis loop in the I–V characteristics\cite{6}, revealing fundamental aspects of its memory effects and their dependence on the timescale of the input signal \cite{7,8,29}. We investigated the I–V response under various voltage sweep rates ($v$), defined as the ratio of a fixed voltage step ($\Delta V=0.02 \ \mathrm{V}$) to a variable step duration ($\Delta t$), i.e. $v=\Delta V/\Delta t$. At intermediate sweep rates, a pronounced hysteresis loop appears in the I-V curve (FIG.~\ref{fig:3}a). This occurs because the internal conductive channel transitions from LCS to HCS during the forward sweep but then remains in the HCS during the reverse sweep due to a slow relaxation time, as detailed in FIG.~\ref{fig:3}b. The hysteresis disappears at extreme sweep rates; at very slow rates, the system has sufficient time to track the voltage changes, causing the I-V paths to overlap, while at very fast rates, the system cannot respond in time, thus remaining in its initial LCS. Consistent with the hopping transport, the hysteresis associated with tunneling transport also shows a significant dependence on the sweep rate, as detailed in  Section 3 of the Supplementary Material.
\subsection{\label{sec:level2d}Emulation of Tunable Synaptic Plasticity  }
A key functionality of our molecular-scale memristor model is its capacity to exhibit complex temporal responses analogous to synaptic plasticity, a behavior that stems directly from its coupled, dual-timescale dynamics.\cite{2,3,31} In biology, this plasticity is the basis of memory and learning, as it relies on history-dependent processes like neurotransmitter activity.\cite{36,39} This parallel is compelling because the underlying physics mirrors the biological processes: the fast electronic transport in our device to resemble rapid action potential propagation, while the slow chemical kinetics correspond to the gradual processes of Ca\textsuperscript{2+} influx and neurotransmitter release.\cite{2,21}

\begin{figure*}
    \centering
    \includegraphics[width=0.9\linewidth]{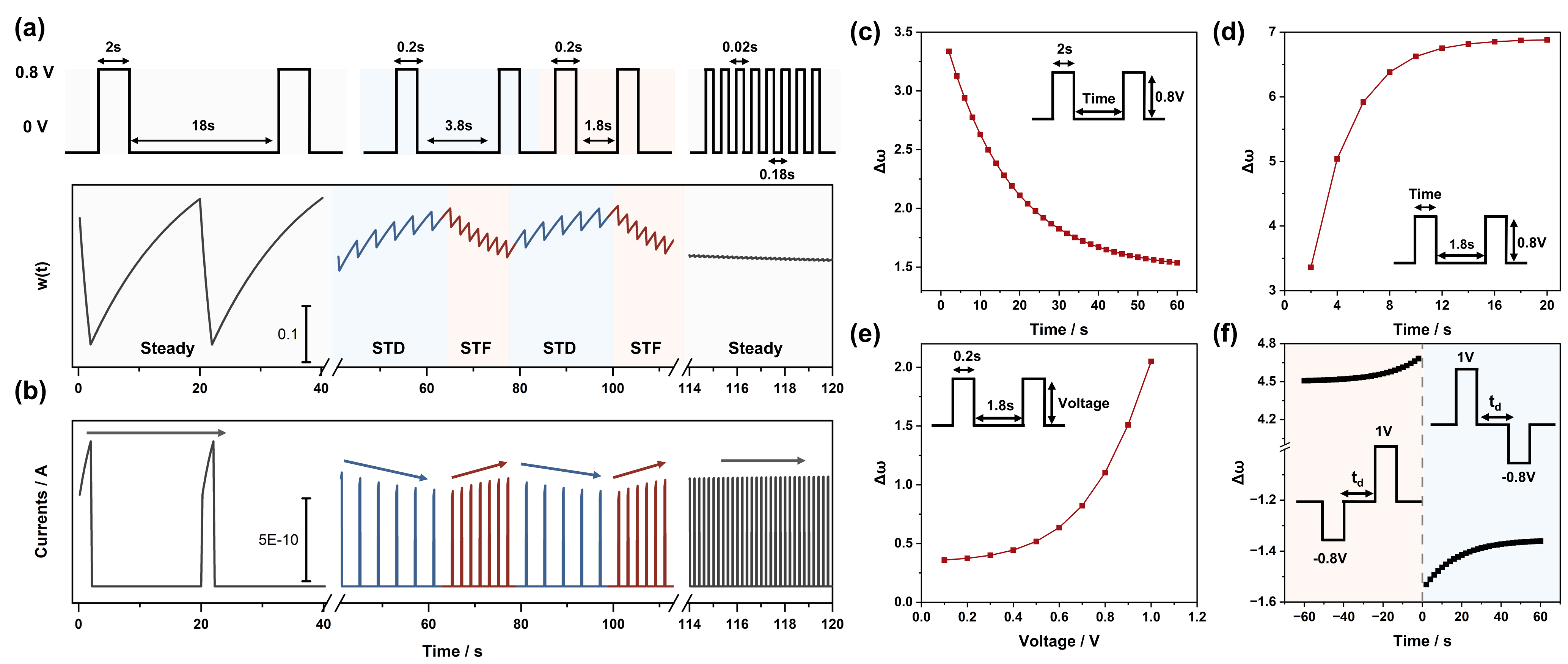}
    \caption{Simulated synaptic plasticity in the hopping-based memristor. (a) Time evolution of LCS fraction ($w(t)$), under low-, moderate-, and high-frequency pulse voltage. (b) Corresponding current traces showing frequency-dependent short-term plasticity, where facilitation and depression emerge most clearly at moderate input frequency. Current change, $\Delta w$ , versus pulse relaxation time (c), pulse duration (d), and pulse amplitude (e). (f) $\Delta w$ as a function of relative spike timing, revealing device-level STDP learning rule. Inset: paired-pulse stimulation protocol.
}
    \label{fig:4}
\end{figure*}

To simulate STP, we applied continuous voltage pulses at varying frequencies to the molecular dynamic memristor, modulating its internal redox state (FIG.~\ref{fig:4}a). This protocol reveals a frequency-dependent plasticity analogous to synaptic behavior (FIG.~\ref{fig:4}b): high-frequency pulses lead to a cumulative conductance enhancement analogous to synaptic facilitation due to insufficient relaxation, while low-frequency pulses result in a conductance suppression analogous to synaptic depression by allowing for full relaxation. Consequently, the device demonstrates effective plasticity only within an intermediate frequency range, with negligible effects at extreme frequencies. Therefore, this plasticity arises only when the input pulse frequency matches the device’s intrinsic relaxation timescale. 

Beyond simply emulating synaptic behavior at specific frequencies, we further demonstrate that this plasticity can be finely tuned by adjusting the pulse parameters. Synaptic modulation ($\Delta w$) is quantified by the relative current change, calculated as $\Delta w={\frac{I_n-I_0}{I_0}}
\label{eq:10}$ , where  is the initial current and is the current after the n-th pulse.\cite{31} This modulation is further dependent on the pulse relaxation time, duration, and amplitude, as illustrated in FIG.~\ref{fig:4}c-e. While longer inter-pulse intervals weaken the facilitation by providing sufficient time for the system to relax back toward the LCS, the effect is enhanced by either a longer pulse duration or a higher amplitude, as both factors more effectively drive the system toward the HCS. We further simulate STDP, a hallmark of long-term synaptic learning.\cite{39} By applying pre- and post-synaptic pulse pairs (the pre-synaptic pulse corresponding to a negative voltage and the post-synaptic pulse to a positive voltage)\cite{16}, the device exhibits both depression ($t_d>0$) and facilitation ($t_d<0$), as shown in FIG.~\ref{fig:4}f. The corresponding results for synaptic plasticity emulation using the tunneling-based memristor are presented in  Section 4 of the Supplementary Material. 

\subsection{\label{sec:level2e}Implementation of Reservoir Computing
}
Reservoir Computing (RC) is a brain-inspired paradigm that leverages the transient dynamics of a fixed, nonlinear system (the "reservoir") to project temporal inputs into a higher-dimensional feature space.\cite{50,51,52} Critically, unlike conventional artificial neural networks (ANNs) that utilize the static, non-volatile states of memristors, RC's performance is fundamentally determined by the quality of a device's intrinsic dynamics and fading memory.\cite{53} This makes RC not just a computing method, but an ideal framework for evaluating the computational potential of our dynamic memristor. In this work, we construct the RC framework using a time-multiplexing strategy,\cite{52} as shown in FIG.~\ref{fig:1}b, with the detailed implementation provided in Section 5 of the Supplementary Material.

Within framework, we probe and harness the device's dynamics by tuning two key hyperparameters:

1. The input frequency,\cite{32,52} which is critically linked to the virtual node interval and the mask length by the relationship:
\begin{equation}
f = \frac{1}{\chi} = \frac{1}{M \times \theta}
\label{eq:11}.
\end{equation}
It must be carefully tuned to the memristor's intrinsic relaxation time. An appropriately chosen timescale effectively probes the memristor’s slow dynamic evolution, which in turn generates the rich and separable state trajectories required for computation. 

2. Bias mapping range ($[V_{\mathrm{Min}}, V_{\mathrm{Max}}]$)\cite{24,54,55} which maps the input signal to the device's operating regime, directly influencing electron occupancy during charge transport and thus governing the overall dynamics. \cite{24,54,55}

We evaluated the computational capabilities using two types of tasks: waveform recognition and chaotic system prediction. 

\subsection{\label{sec:level2f}Dependence of RC Performance on Frequency
}
\begin{figure*}
    \centering
    \includegraphics[width=0.75\linewidth]{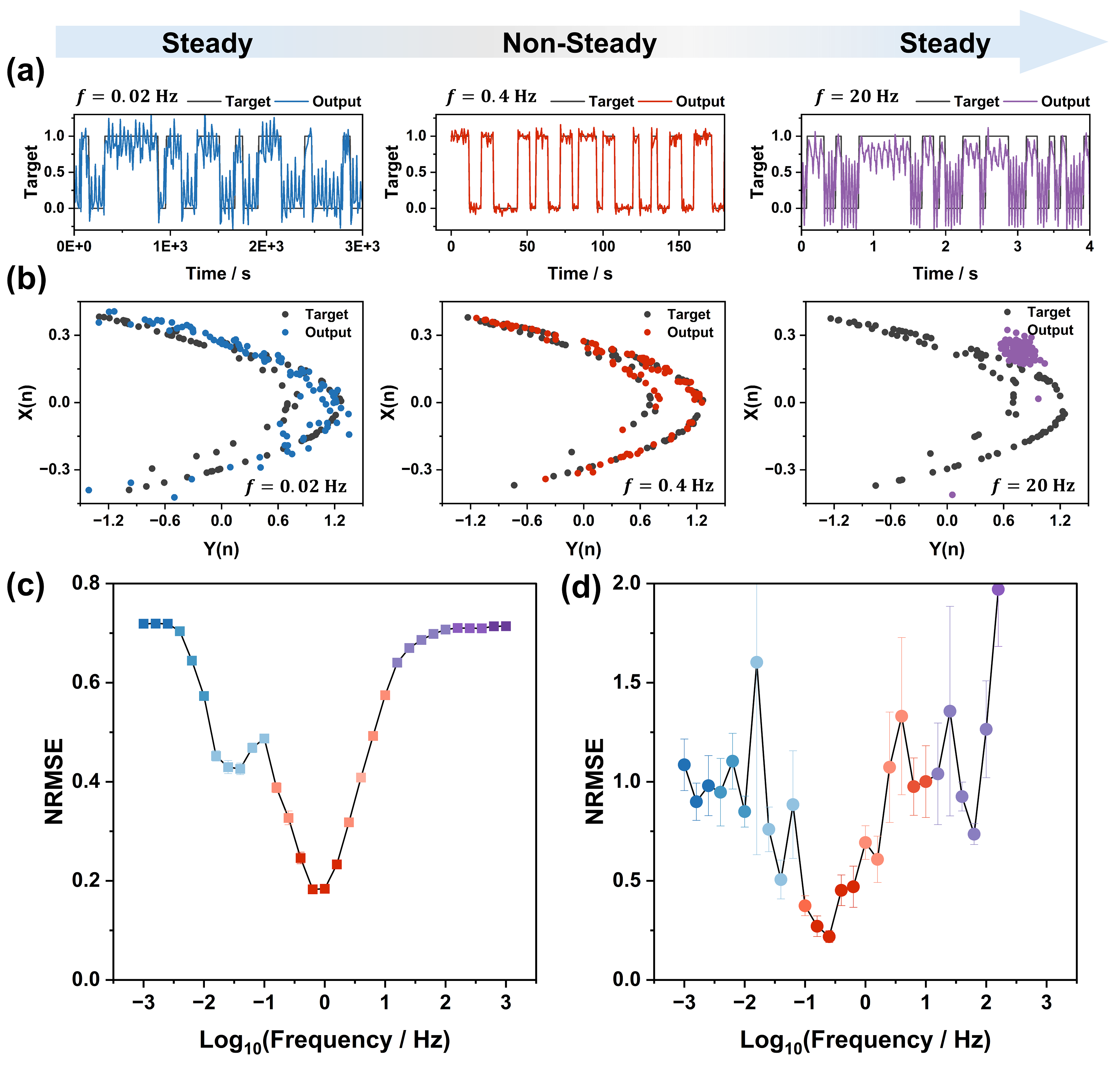}
    \caption{Reservoir computing performance of the hopping-based memristor under different dynamic conditions. Operation at moderate input frequencies drives the device into a non-steady-state regime, enabling effective information processing. Representative demonstrations of single-run task performance: (a) Waveform classification: The gray line represents the target output (‘0’ for sine, ‘1’ for square), while the colored lines show the predictions. (b) Hénon map prediction task: a two-dimensional plot of the intended targets (as gray points) with the RC-generated outputs (in other colors). The NRMSE statistics for (c) waveform classification and (d) Hénon map prediction are plotted as functions of input frequency. Different colors indicate different dynamic conditions.
}
    \label{fig:5}
\end{figure*}

We begin with the waveform recognition task. The bias mapping range was set to [–0.5 V, 1.5 V]. We set the mask length $M$ to =5 and used eight parallel memristors, a configuration chosen to balance computational efficiency and state richness.\cite{24} A mixed sequence of square and sine waves was generated, with square waves labeled as 1 and sine waves as 0. After training the output layer via linear regression, the recognition results are shown in FIG.~\ref{fig:5}a. Under non-steady-state conditions, the system achieved accurate classification with a normalized root mean square error (NRMSE)\cite{56} approximately 0.15 and a success rate of 100\%. In contrast, under steady-state conditions, the system failed to distinguish between waveforms, highlighting the critical role of dynamic behavior in enabling effective information processing.

The Hénon map is a typical discrete-time dynamical system exhibiting chaotic behavior. It defines a nonlinear two-dimensional mapping that transforms a point $(x(n),y(n))$ on the plane into a new point $(x(n+1),y(n+1))$, as described by:
\begin{equation}
\begin{aligned}
x(n+1) &= y(n) - 1.4x(n)^2, \\
y(n+1) &= 0.3x(n) + w(n).
\end{aligned}
\label{eq:12}
\end{equation}
Here,  $w(n)$is Gaussian noise with zero mean and a standard deviation of 0.05.\cite{57} The task aims to learn the underlying nonlinear mapping and reconstruct the chaotic attractor. With the mask length set to $M$ = 5 and 25 memristor devices used in parallel, step-by-step prediction was employed by feeding each predicted output back as the input for the next step. In a non-steady-state regime, the system accurately predicts up to five steps ahead, achieving a NRMSE of approximately 0.2. Although prediction errors accumulate over time, the system still successfully reconstructs the attractor over a 100-step prediction horizon, maintaining its overall shape with only minor phase shift. This demonstrates the device’s strong memory capability and robustness in capturing the trajectory of chaotic systems.\cite{25} In contrast, the steady-state system fails to reconstruct the attractor, indicating a clear limitation in capturing the characteristics of chaotic dynamics. The attractor reconstruction is illustrated in FIG.~\ref{fig:5}b, while the one-dimensional time-series prediction is presented in FIG. S5.

To further investigate the relationship between input frequency and information processing capability, we plotted the NRMSE for both tasks as a function of input frequency, as shown in FIG.~\ref{fig:5}c, d. These plots illustrate how performance varies with signal frequency. To ensure reliability and minimize the influence of randomness, each frequency point was evaluated over 50 independent runs, and the average result is reported with error bars. The results indicate that, regardless of the task type, optimal computational performance is achieved only when the input frequency falls within a specific range that places the system in a non-steady state. Combined with earlier analysis, this optimal frequency corresponds to the sweep rate at which non-steady-state features are observed experimentally. Based on this, we translate the characteristic sweep rate into the virtual node time interval $\theta$ in reservoir computing (i.e., the input frequency, as defined in Eq.~\eqref{eq:11}), thereby establishing a direct link between the system’s dynamic characteristics and its computational performance, as described by:
\begin{equation}
\theta = \frac{\Delta V}{v} \cdot c_n
\label{eq:13}.
\end{equation}
$\Delta V$ is the voltage step used in the I–V simulation, set to 0.02 V as commonly used in experiments. $c_n$ is a fitting constant, which can be adjusted for different molecular systems, and the expression presented here serves as a general methodological guideline. For the tunneling-based memristor, the dependence of its information processing capabilities on the timescale is analyzed in Section 6 of the Supplementary Material.

\begin{figure}[!htbp]
    \centering
    \includegraphics[width=1\linewidth]{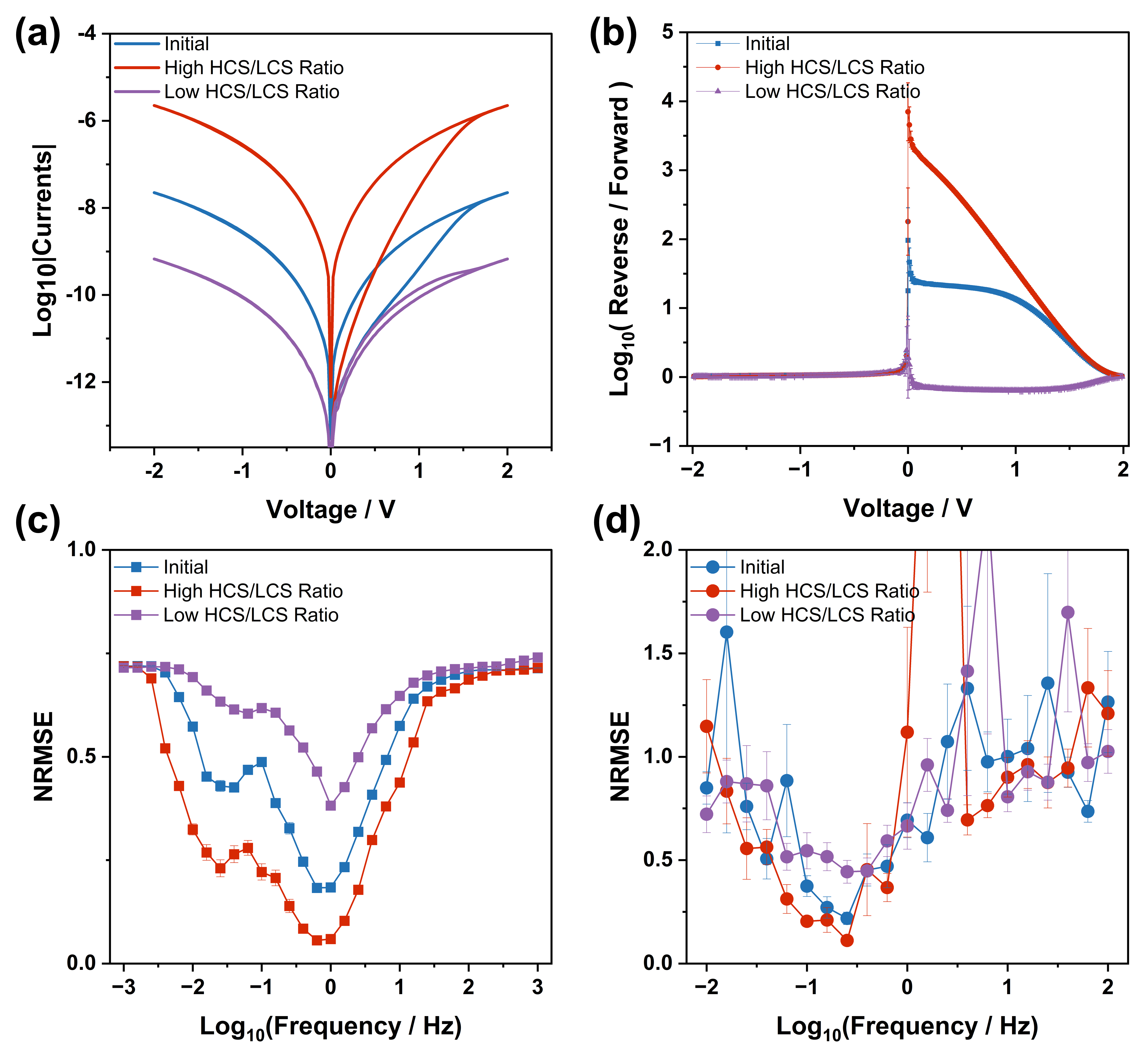}
    \caption{Effect of the HCS/LCS ratio on hysteresis and computational performance of the hopping-based memristor. (a) Dynamic current responses; (b) hysteresis degree for different HCS/LCS ratios. (c) Waveform classification task; (d) Hénon map prediction task: NRMSE versus input frequency for different HCS/LCS ratios.
}
    \label{fig:6}
\end{figure}

Building on our finding that maximal I-V hysteresis at an optimal timescale yields the best performance, we next investigated if the intrinsic magnitude of this hysteresis, governed by the HCS/LCS ratio, could further enhance computational capability. A larger HCS/LCS ratio, achieved by computationally increasing the HCS 100-fold while holding the LCS constant, yields two synergistic benefits. First, it produces a more pronounced hysteresis loop, reflecting a more significant apparent non-steady-state behavior. Second, it increases the overall current magnitude, which enhances the system's robustness against noise (FIG.~\ref{fig:6}a, b). The superior computational performance from a large HCS/LCS ratio is evidenced by a broader effective frequency range and a lower NRMSE. In contrast, a smaller ratio diminishes this capability, resulting in a narrower frequency range and increased error (as demonstrated in FIG.~\ref{fig:6}c, d). However, the performance enhancement offered by a large HCS/LCS ratio is highly task dependent. For applications like waveform classification, which rely on a clear distinction between discrete states to classify binary inputs, a high ratio is critical and yields significant improvement. In contrast, for tasks strongly dependent on temporal features like Hénon map prediction, the computation leverages the rich internal dynamics of the memristor, making high state separability less crucial and thus limiting the performance enhancement from a large HCS/LCS ratio. Furthermore, this optimization is subject to a physical limit: at very high currents, the dominance of the HCS renders further increases in the HCS/LCS ratio ineffective for enhancing hysteresis. The corresponding analysis for the tunneling transport is provided for completeness in Supplementary Material FIG. S7. We also discuss the effect of increasing the current magnitude at a constant HCS/LCS ratio in FIG. S8.

\subsection{\label{sec:level2g}Dependence of RC Performance on Bias Range}
We now turn to the second key hyperparameter governing RC performance: the bias mapping range. In time-multiplexed RC, a one-dimensional physical node is “unfolded” into many virtual nodes by feeding it a continuous input stream that is multiplied by a random mask and then mapped onto a chosen voltage window $[V_{\mathrm{Min}}, V_{\mathrm{Max}}]$. The random mask enriches the dynamics by forcing the device to sample different input amplitudes within each mask period.\cite{24,52} While, as indicated by Eqs.~\eqref{eq:5}--\eqref{eq:8} the rate constant of the system's internal state evolution is bias-dependent, the choice of the bias mapping range directly influences the system's non-steady-state dynamics during RC computation. Consequently, based on our earlier discussion of the interplay between input frequency and system dynamics, the bias mapping range, together with the input frequency, jointly determines the diversity of internal states — a critical factor that ultimately affects RC performance. 

\begin{figure*}
    \centering
    \includegraphics[width=0.9\linewidth]{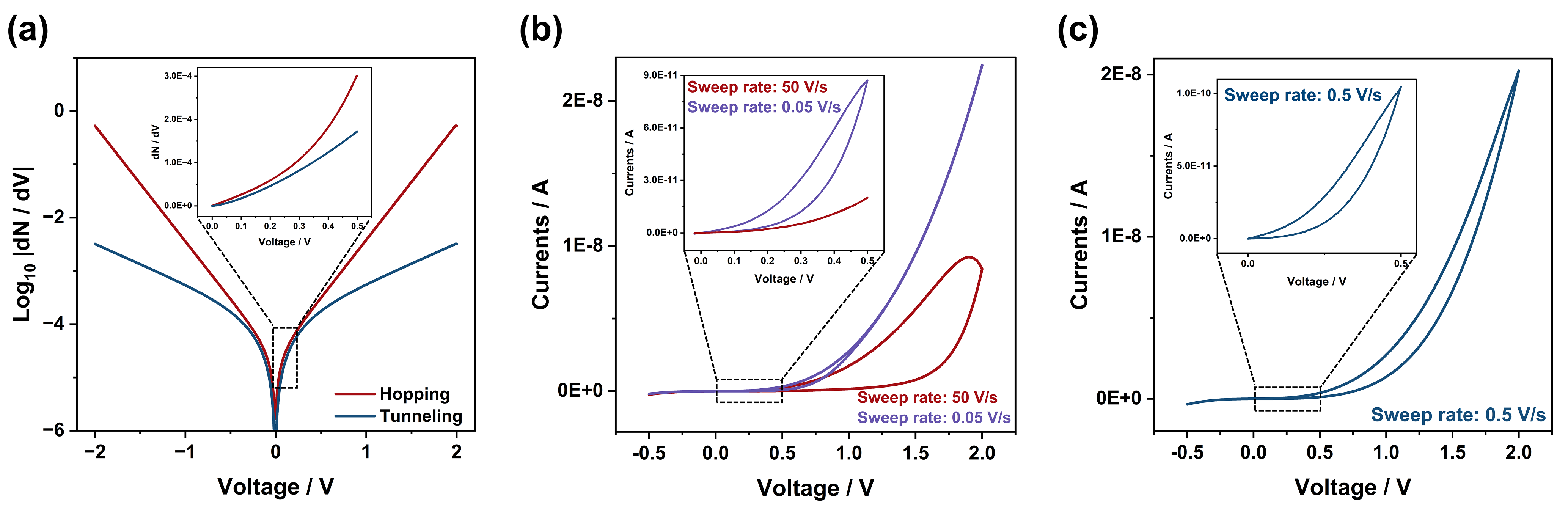}
    \caption{Bias-dependent electron occupancy ($N$) variations and I-V characteristics under different bias ranges for hopping- and tunneling-based memristors. (a) $ {dN}/{dV} $ on a logarithmic axis for both mechanisms, reflecting the sensitivity of $N$ to the applied bias; inset: the same data on a linear scale within [0, 0.5 V], showing the stronger non-linearity of the hopping transport. (b) Hopping transport: I-V characteristics plotted in [-0.5 V, 2 V] range at high (50 V/s) and low (0.05 V/s) sweep rates; a pronounced hysteresis loop appears only at the high rate. Inset: within narrow [0, 0.5 V] bias window, hysteresis emerges only at the low rate. (c) Tunneling transport: I-V characteristics at 0.5 V/s sweep rate; a consistent hysteretic loop is observed over both the wide range ([-0.5 V, 2 V], main panel) and the narrow low-bias window ([0, 0.5 V], inset).}
    \label{fig:7}
\end{figure*}

The hopping and tunneling transports exhibit fundamentally different dependencies of redox state occupancy on the applied bias. As shown in FIG.~\ref{fig:7}a, the  in the hopping transport increases exponentially with bias, spanning a wide dynamic range. In contrast, the tunneling transport takes on a more gradual linear-to-exponential evolution with a narrower range. This disparity has practical consequences. To capture the dynamics of the hopping-based memristor at high bias we must sweep the voltage rapidly, yet those same fast sweeps outrun the slower low-bias kinetics and suppress the associated memory effects. Different bias windows therefore demand different sweep rates (FIG.~\ref{fig:7}b). By contrast, the tunneling-based memristor can be characterized with a single sweep rate across the full bias range, because its kinetics remain slow enough for the system to follow the stimulus at all voltage range (FIG.~\ref{fig:7}c).

\begin{figure}
            \centering
            \includegraphics[width=1\linewidth]{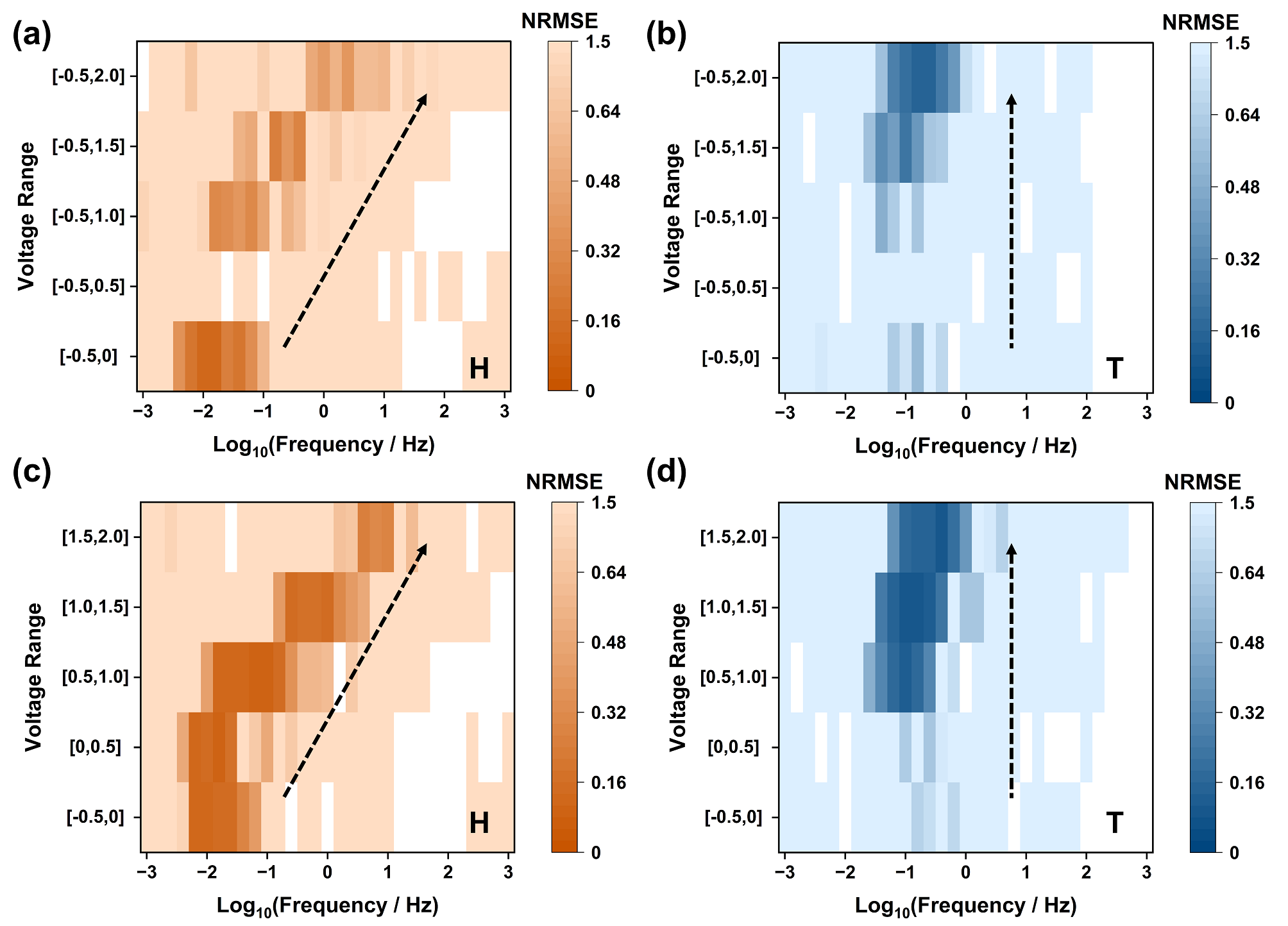}
            \caption{Hénon map prediction performance of the hopping-based and tunneling-based memristor under various bias voltage conditions. (a) Hopping; (b) Tunneling: NRMSE versus input frequency for different bias window widths, with keeping the low voltage bound at -0.5 V. (c) Hopping; (d) Tunneling: NRMSE versus input frequency at a fixed bias window of 0.5 V, compared across different bias ranges. White regions indicate NRMSE > 1.5.}
            \label{fig:8}
        \end{figure}

We then investigated the system's information processing capability for the Hénon map prediction task across various bias ranges. Specifically, the minimum bias was fixed at -0.5 V, while the maximum bias was progressively increased from 0 V to 2 V with 0.5 V increments. The results, shown in FIG.~\ref{fig:8}a, b, reveal that the two types of transport exhibit distinct performance trends under these varying ranges. The optimal input frequency for the hopping transport increases with the bias range to accommodate its shifting dynamics, whereas the tunneling transport maintains a consistent optimal frequency, as its full dynamics can be captured at a fixed sweep rate. 

These dynamic differences are directly translated into computational power. The hopping-based memristor performs best in narrow windows, where its full non-linear dynamics are visible. In contrast, the tunneling-based memristor improves as the window widens: although its dynamics are always accessible, a broader range activates additional redox intermediates and enriches the reservoir state space. To validate these findings, we fixed the bias window of the hopping transport to 0.5 V and slid it across the bias axis. As shown in FIG.~\ref{fig:8}c, the accuracy remained high everywhere, indicating that the dynamic features are fully captured within each window. For the tunneling transport, however, performance varies significantly with the position of the bias window (FIG.~\ref{fig:8}d). At low biases, limited nonlinearity results in weaker predictive performance, whereas high-bias windows offer stronger non-linearity and deeper state richness, boosting accuracy. 

\begin{figure}[b]
            \centering
            \includegraphics[width=1\linewidth]{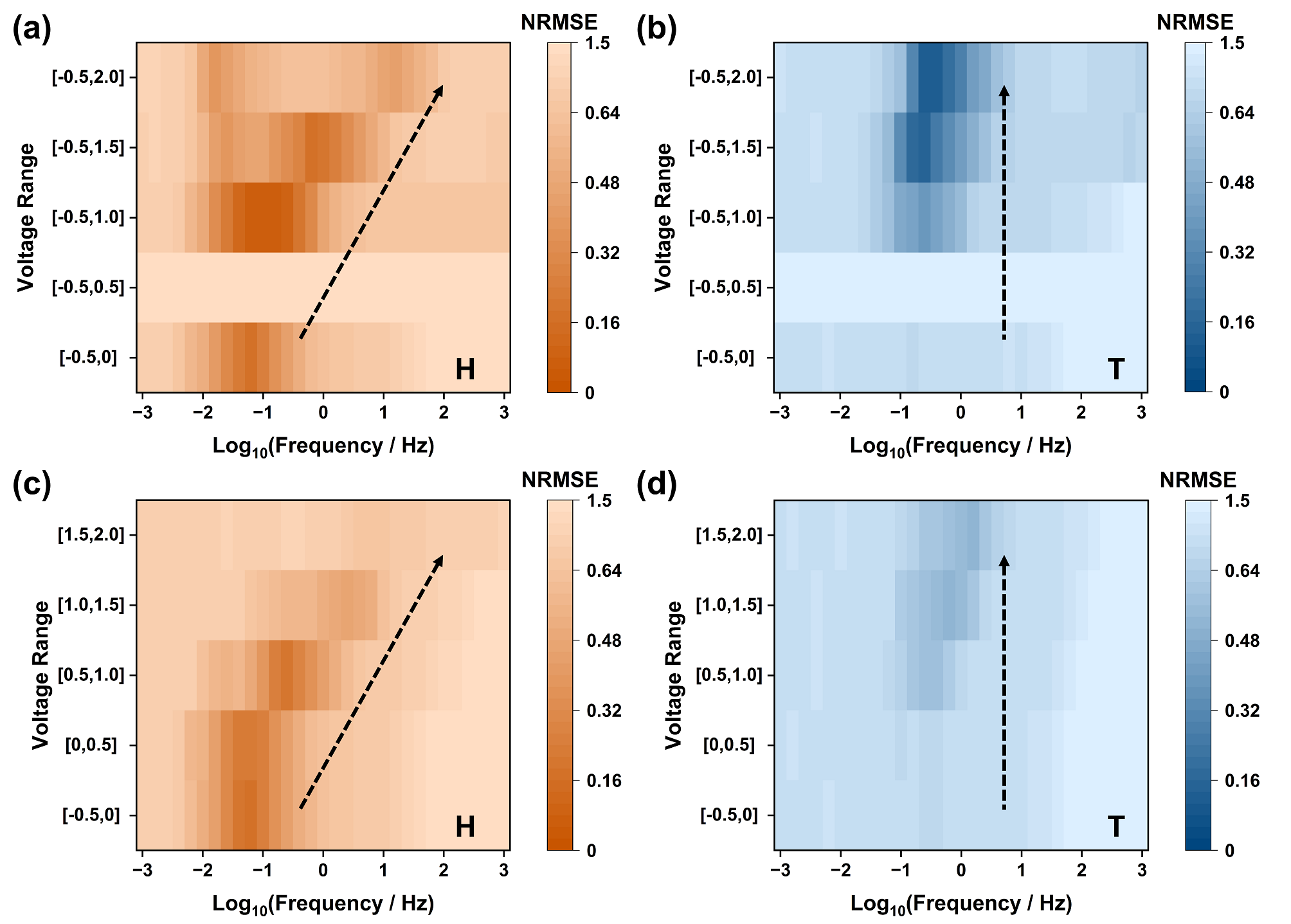}
            \caption{Waveform classification performance of the hopping-based and tunneling-based memristor under various bias voltage conditions. (a) Hopping; (b) Tunneling: NRMSE versus input frequency for different bias window widths, with keeping the low voltage bound at -0.5 V. (c) Hopping; (d) Tunneling: NRMSE versus input frequency at a fixed bias window of 0.5 V, compared across different bias ranges.}
            \label{fig:9}
        \end{figure}

A parallel analysis for the waveform classification task leads to the same qualitative picture. As analyzed above, waveform classification relies heavily on pronounced hysteresis to distinguish between states. So, the hopping-based memristor must balance dynamic coverage against hysteresis width. This balance is achieved in the [–0.5 V, 1 V] bias range (FIG.~\ref{fig:9}a), as narrower windows shrink hysteresis, while wider ones hide part of the dynamics. The tunneling-based memristor, in contrast, is more straightforward: both its hysteresis and its classification accuracy peak in the widest window tested (FIG.~\ref{fig:9}b). Consequently, for both types of transport, reducing the bias range from their respective optima degrades performance due to weakened hysteresis (FIG.~\ref{fig:9}c, d). By contrast, tasks such as chaotic time-series prediction are characterized by a greater dependence on high-dimensional state projection than on hysteresis per se.

In summary, three guidelines emerge for choosing a bias window: (1) the device’s full dynamics must be accessible, (2) those dynamics should exhibit strong non-linearity, and (3) the window should be tailored to the specific computational task. Finally, because the device responds symmetrically to positive and negative bias, a symmetric voltage window (e.g., ±1 V) with the same dynamics in both directions will reduce the reservoir’s state diversity and harming performance and so adopting an asymmetric bias configuration becomes essential.

\section{\label{sec:level1d}conclusion}
In this work we advanced molecule-based neuromorphic electronics from proof-of-concept demonstrations toward a principled design discipline. By unifying Landauer- and Marcus-type electron transport with slow chemical dynamics, we established a general dynamic memristor model that quantitatively predicts both device physics (conductance hysteresis, STP, STDP) and system-level behavior when embedded in a reservoir-computing (RC) architecture. The model reveals an experimentally actionable design rule that RC performance peaks when the input frequency and bias mapping window are commensurate with the molecule’s intrinsic reaction rates, thereby keeping the device in a richly non-steady regime that maximally diversifies its internal states.

The model suggests explicit relationships between molecular kinetics and computational hyperparameters, which could enable researchers to begin with a target task, derive the required kinetic window, and design molecules and device environments to match. We propose this framework as a potential roadmap for task-driven fabrication of memristive junctions, guiding selections of redox cores, substituents, electrolytes, and electrode chemistry. Furthermore, the formalism may extend from single-molecule junctions to self-assembled monolayers and hybrid organic–inorganic stacks, facilitating large-area arrays where sensing, memory, and processing coexist within the molecular layer. By connecting computational performance with molecular timescales, we hope that this study invites further exploration for chemistry-driven innovation in post-Moore architectures.

\section*{SUPPLEMENTARY MATERIAL}

A comparison of model simulation to actual results (Figure S1); I-V characteristics, synaptic plasticity, and the analysis of frequency dependence of reservoir computing performance for the tunneling-based memristor (Figures S2-S5); computational performance under linear current (Figure S6); the effect of the HCS/LCS ratio in the tunneling-based memristor (Figure S7); and the impact of current magnitude on the performance of two types of transport (Figure S8).

\section*{ACKNOWLEDGMENTS}

This work is supported by the ISF-NSFC Joint Scientific Research Program (22361142833), the Open Project of the State Key Laboratory of Supramolecular Structure and Materials (sklssm202527), and ‘the Fundamental Research Funds for the Central Universities’.

\section*{AUTHOR DECLARATIONS}

\noindent \textbf{Conflict of Interest}

\noindent The authors have no conflicts to disclose.

\noindent \textbf{Author Contributions}

\noindent \textbf{Yueqi Chen}: Conceptualization (equal); Methodology (equal); Formal analysis (equal); Investigation (equal); Writing – original draft (equal). \textbf{Xuan Ji}: Conceptualization (equal); Formal analysis (equal); Investigation (equal); Writing – review \& editing (equal). \textbf{Xi Yu}: Conceptualization (equal); Supervision; Writing – review \& editing (equal).

\section*{Data Availability Statement}

The data that support the findings of this study are available from the corresponding authors upon reasonable request.

\section*{References}
\nocite{*}
\bibliographystyle{aipnum4-1}   
\bibliography{reference}

\end{document}